\documentclass[11pt]{article}

\usepackage{amssymb,amsmath,amsthm,mathrsfs,latexsym,amsxtra,graphicx,appendix,epstopdf,feynmf,subfig,hyperref,setspace,fix-cm,color,cite,hyperref}
\usepackage{verbatim}
\usepackage[a4paper,left=2cm,right=1cm,top=4cm,bottom=4cm,bindingoffset=5mm]{geometry}
\usepackage{pgfplots}
\pgfplotsset{compat=newest}
\usepackage{mathtools}
\usepackage{tensor}
\usepackage{siunitx}
\usepackage{tikz}
\usetikzlibrary{trees}
\usetikzlibrary{decorations.pathmorphing}
\usetikzlibrary{decorations.markings}
\usetikzlibrary{decorations.markings}
\usepackage{float}
\usepackage{rotating}
\usepackage{simplewick}
\usepackage{cancel}
\usepackage{ marvosym }
\usepgfplotslibrary{fillbetween}
\usetikzlibrary{patterns}

\usepackage{mathtools}

\usepackage{setspace}
\usepackage{lipsum}
 \usepackage{multirow}
\usepackage{fullpage}
\usepackage{amsmath}
\usepackage{amssymb}
\usepackage{setspace}
\usepackage{bbm}
\usepackage{dsfont}
\usepackage{graphics}
\usepackage{longtable}
\usepackage[font=footnotesize,labelfont=bf,justification=centerlast,width=.9\textwidth]{caption}
\usepackage{color}
\usepackage{etoolbox}
 \usepackage{multirow}
\usepackage{booktabs}
\usepackage{array}
\usepackage{hyperref}
\usepackage{cite}
\usepackage[normalem]{ulem}
\hypersetup{
    bookmarks=true,%
    colorlinks,%
    citecolor=blue,%
    filecolor=blue,%
    linkcolor=blue,%
    urlcolor=blue
}

\newcommand{\twobytwo}[4]{\left(\begin{array}{cc}#1&#2\\#3&#4\end{array}\right)}
\newcommand\rref[1]{(\ref{#1})}
\newcommand{\be}{\begin{equation}}
\newcommand{\ee}{\end{equation}}
\newcommand{\bes}{\begin{equation*}}
\newcommand{\ees}{\end{equation*}}
\newcommand{\bea}{\begin{eqnarray}}
\newcommand{\eea}{\end{eqnarray}}
\newcommand{\beas}{\begin{eqnarray*}}
\newcommand{\eeas}{\end{eqnarray*}}

\newcommand{\bmat}{\begin{bmatrix}}
\newcommand{\emat}{\end{bmatrix}}

\def\le{\left}
\def\ri{\right}

\def\z{\mathfrak{z}}

\newcommand{\CC}{\mathbb{C}} 

\newcommand{\QQ}{\mathbb{Q}} 
 
\newcommand{\ZZ}{\mathbb{Z}}
\newcommand{\NN}{\mathbb{N}}

\newcommand{\Tr}{{\rm {Tr}}}

\renewcommand{\H}{\mathbb{H}}
\newcommand{\Z}{\mathbb{Z}}

\newcommand{\ie}{{\it i.e.~}}

\newcommand{\fr}{{f_{\textrm{R}}}}

\newtheorem{thm}{Theorem}[section]
\newtheorem{defn}{Definition}[section]

\begin{document}
\numberwithin{equation}{section}
{
\begin{titlepage}

\hfill { \large CERN-TH-2019-163 }

\begin{center}

\hfill \\
\hfill \\
\vskip 0.75in

{\Large \bf Siegel Paramodular Forms from Exponential Lifts:\\ 
	Slow versus Fast Growth}\\

\vskip 0.4in

{\large Alexandre Belin${}^a$, Alejandra Castro${}^b$, Christoph A.~Keller${}^{c}$, and  Beatrix M\"uhlmann${}^b$}\\
\vskip 0.3in

${}^{a}${\it CERN, Theory Division, 1 Esplanade des Particules, Geneva 23, CH-1211, Switzerland}
\vskip .5mm
${}^{b}${\it Institute for Theoretical Physics, University of Amsterdam,
Science Park 904, Postbus 94485, 1090 GL Amsterdam, The Netherlands} \vskip .5mm

${}^{c}${\it Department of Mathematics, University of Arizona, Tucson, AZ 85721-0089, USA} \vskip .5mm

\texttt{a.belin@cern.ch, a.castro@uva.nl, cakeller@math.arizona.edu, b.muhlmann@uva.nl}

\end{center}

\vskip 0.35in

\begin{center} {\bf ABSTRACT } \end{center}
We investigate the growth of Fourier coefficients of Siegel paramodular forms built by exponentially lifting weak Jacobi forms, focusing on terms with large negative discriminant. 
To this end we implement a method based on deforming contours that expresses the coefficients of all such terms as residues. We find that there are two types of weak Jacobi forms, leading to two different growth behaviors: the more common type leads to fast, exponential growth, whereas
a second type leads to slower growth, akin to the growth seen in ratios of theta functions. We give a simple criterion to distinguish between the two types, and give a simple closed form expression for the coefficients in the slow growing case.
In a companion article \cite{Belin:2019rba},
we provide physical applications of these results to  symmetric product orbifolds and holography.

\vfill

\noindent \today

\end{titlepage}
}

\newpage

\tableofcontents

\newpage

\section{Introduction}

There is a long standing mathematical connection between Jacobi forms \cite{MR781735} and Siegel modular forms (SMF) \cite{MR2385372}. In fact one of the original motivations for studying Jacobi forms \cite{MR781735} came from the Fourier-Jacobi expansion of SMF. Another connection is given by the exponential lift \cite{MR1616929}: A weak Jacobi  form (wJf) $\varphi$ of weight 0 and index $t$ can be lifted to a Siegel paramodular form as in 
\be\label{introexplift}
\Phi_k(\Omega)=\textrm{Exp-Lift}(\varphi)(\Omega)= q^A y^B p^C \prod_{\substack{n,l,r\in\ZZ
		\\(n,l,r)>0}} (1-q^n y^l p^{tr})^{c(nr,l)}~.
\ee
Here the $c(n,l)$ are the Fourier coefficients of  $\varphi$, and
the notation used on the right hand side will be explained in the next section.
The best known example of this is the Igusa cusp form $\chi_{10}$ \cite{Igusa35,Igusa}, which is the lift of the wJf $2\phi_{0,1}$. In fact, $\chi_{10}$ is one of the generators of the ring of (holomorphic) SMF.

In physics $\chi_{10}$ plays a crucial role for the counting of black hole entropy \cite{Dijkgraaf:1996it}. Physicists tend to be more interested in \emph{meromorphic} SMF, as they are interested in exponentially growing Fourier coefficients: the object investigated in \cite{Dijkgraaf:1996it} is thus $1/\chi_{10}$. Initially, only terms of very large positive discriminant were investigated. Later on, much more detailed information was obtained also for other terms \cite{Sen:2007qy,Sen:2011mh} --- see in particular \cite{Dabholkar:2012nd} for a mathematical treatment of this.
In all this it is important that the meromorphic SMF is an exponential lift: it can then be interpreted as coming from what is called a symmetric orbifold. It is thus natural to consider exponential lifts other than just $1/\chi_{10}$, a program we started in \cite{Belin:2016knb,Belin:2018oza}.
In this article we continue to analyze the coefficients of Siegel (para-)modular forms that come from exponential lifts of a wJf.

To set up our notation, consider the Fourier expansion of
\be\label{eq:introPhik}
\frac1{\Phi_k(\Omega)} = \sum d(m,n,l) p^m q^n y^l\ .
\ee
We are interested in the growth behavior of the coefficients $d(m,n,l)$. That behavior 
greatly depends on the \emph{discriminant} $\Delta = 4mn-l^2$. If $\Delta$ is large and positive, then the behavior of $d(m,n,l)$ is universal and mostly independent of the underlying wJf $\varphi$, as was worked out in \cite{Belin:2016knb}. 
For negative discriminant, the situation is more interesting: As was found in \cite{Benjamin:2015vkc} and \cite{Belin:2018oza}, the growth of $d(m,n,l)$ depends on the specific choice of the underlying wJf $\varphi$. In particular, for certain choices of $\varphi$, it can be much slower than generically expected. 

In this article we make this distinction more precise. For terms with very large negative discriminants we find a dichotomy: $d(m,n,l)$ is either \emph{fast growing}, which means roughly that $\log |d(m,n,l)|$ grows linearly, or it is \emph{slow growing}, which means that $\log |d(m,n,l)|$ grows like a square root. More precisely, for slow growing forms the generating function of the coefficients is a ratio of theta functions, whereas for fast growing forms, it is a product with exponentially growing exponents.

To establish this, we first describe a method pioneered   in \cite{Sen:2011mh} for extracting the Fourier coefficients of negative discriminant terms in (\ref{eq:introPhik}). It relies on deforming the contour of the integral that extracts the sought-after Fourier coefficient from (\ref{eq:introPhik}) to a contour that gives manifestly zero. In the process however, due to the fact that $1/\Phi_k$ is meromorphic, one crosses one or more poles, and therefore picks up the residues of those poles. (In physics this is called \emph{wall-crossing} \cite{Cheng:2007ch}.) The Fourier coefficient is thus given by a sum of residues. We used this procedure in \cite{Belin:2018oza} to compute the coefficients of exponential lifts that only contained certain classes of poles. In the present article, we explain how to apply it for a general class of poles appearing in $1/\Phi_k$.

It turns out that the more negative the discriminant is, the fewer poles need to be taken into account. Conversely, for positive discriminant terms the procedure breaks down. We establish that for terms with close to maximum negative discriminant, only the residue of a single pole contributes: this is what we call the \emph{single pole regime} for the parameters $m,n,l$. For such terms it is particularly simple to give a closed form for the residue, and therefore for $d(m,n,l)$. The terms in the single pole regime are the ones that differentiate between the fast or slow growth of $d(m,n,l)$. To be more precise,
the formula for the Fourier coefficients in this regime is
\be
d(m,n,l)=\int d\tau dz \, {\cal R}_0(\tau,z) q^{-n} y^{-l-mb/t}~. 
\ee
Here ${\cal R}_0$ is the residue of the single pole that contributes, which turns out to be
\be \label{eq:RRR1}
{\cal R}_0(\tau,z) =  (-1)^{2B}q^{-A} y^{B+{b\over t}C} \prod_{l>0} (1-y^l)^{-c(0,l)}   \prod_{\substack{n\geq 0, L\in\ZZ\\(n,L)\neq(0,0)}}(1-q^{n} y^{L})^{-\fr(n,L)}\ ,
\ee
where
\be\label{introfdeff}
\fr(n,L) := \sum_{r=0}^{\infty} c(nr, L-br)~.
\ee
The growth of the $d(m,n,l)$ is therefore completely determined by the growth of the $\fr(n,L)$. Using modular transformation properties, we establish that the $\fr(n,L)$ can only have two possible behaviors, depending on the choice of $\varphi$: They either grow exponentially, or they are bounded.  In the former case, since the $\fr(n,L)$ grow exponentially, we expect  $\log |d(m,n,l)|$ to grow linearly: this is the fast growing case. In the latter case, the generating function (\ref{eq:RRR1}) is then essentially a ratio of $\theta$-type functions, so that $\log |d(m,n,l)|$ grows like a square root: this is the slow growing case. In fact we are able to give simple closed form expressions for the $\fr(n,L)$ in that case.

The organization of this paper is as follows. In Sec.\,\ref{app:smf} we gather the main properties of Siegel (para-)modular forms and exponential lifts that we will use in the subsequent sections; the emphasis is on the meromorphic properties of $1/\Phi_k$. In Sec.\,\ref{sec:wall} we discuss how to extract exactly the Fourier coefficients $d(m,n,l)$ for negative values of the discriminant, via a wall crossing method. Our technique applies to the exponential lift of any Jacobi form, and we present one example in Sec.\,\ref{sec:example}. In Sec.\,\ref{sec:single} we specialize to exponential lifts of a wJf, where the wJf has maximal polarity $\Delta=-b_0^2$.  For these exponential lifts,  we identify the coefficients that are controlled by only one pole crossing; these coefficients define the so-called single pole regime. In Sec.\,\ref{s:Bad} we show how to characterize the residue of the single pole regime with minimal input.  Based on the asymptotic behaviour of the Fourier coefficients in the single pole regime, in Sec.\,\ref{sec:example-slow-fast} we provide a quick method to determine if the exponential lift falls within the fast growth or slow growth class and we provide complementary material in the appendices. In a companion paper \cite{Belin:2019rba}, we discuss the application of these results to AdS/CFT and the landscape of symmetric orbifold CFTs. 



\section{Siegel paramodular forms and exponential lifts}\label{app:smf}

Our focus here will be on paramodular forms built out of an \emph{exponential lift} of a wJf of weight 0 and index $t$. We discuss such forms in Appendix~\ref{app:wjf}. Their exponential lifts are then described by Theorem 2.1 
of \cite{MR1616929}, which states:
\begin{thm}\label{mainthm}
	Let $\varphi\in J^{nh}_{0,t}$ be a nearly holomorphic Jacobi form of weight 0 and index $t$
	with integral coefficients
	\be\label{eq:nhJf}
	\varphi(\tau,z)= \sum_{n,l} c(n,l)q^n y^l\ .
	\ee
	Define\footnote{Although this theorem applies to nearly holomorphic forms, we will only use it for wJf. Note that for a wJf we have $C=tA$.}
	\be\label{eq:abc}
	A = \frac{1}{24}\sum_{l\in\mathbb{Z}} c(0,l)~,\qquad  B = \frac{1}{2}\sum_{l>0} l c(0,l)~, \qquad C = \frac{1}{4}\sum_{l\in\mathbb{Z}} l^2 c(0,l)~,
	\ee
	and 
	\be\label{eq:k1}
	k={1\over 2} c(0,0)~.
	\ee
	Then the exponential lift $\Phi$ of $\varphi$ is the product
	\be\label{explift}
	\Phi_k = \textrm{Exp-Lift}(\varphi)(\Omega)= q^A y^B p^C \prod_{\substack{n,l,r\in\ZZ
			\\(n,l,r)>0}} (1-q^n y^l p^{tr})^{c(nr,l)}\ ,
	\ee
	where $(n,l,r)>0$ means $r >0 \lor (r=0 \land n>0) \lor (n=r=0 \land l <0)$. $\Phi_k$ is
	a meromorphic modular form of weight $k$ with respect to the paramodular group $\Gamma^+_t$ defined below.
	It has a character (or a multiplier system if the weight is half-integral) induced by $v^{24A}_\eta \times v^{2B}_H$.
	Here $v_\eta$ is a 24th root of unity, and $v_H=\pm1$.
\end{thm}

Let us now explain what the properties of $\Phi_k$ are. We use the notation
\begin{align}
p=e^{2\pi i \rho}~,\quad q=e^{2\pi i \tau}~,\quad y=e^{2\pi i z}~,
\end{align}
and it is also convenient to introduce the matrix
\be\label{Omegadef}
\Omega = \left(\begin{array}{cc}\tau&z\\z&\rho\end{array}\right)~.
\ee
With these variables, the Siegel upper half plane $\H_2$ is given by
\be\label{eq:uhp}
\det(\text{Im}(\Omega)) > 0 \ , \qquad \Tr (\text{Im}(\Omega)) > 0\ .
\ee
Our lift $1/\Phi_k(\Omega)$ is a meromorphic function on $\H_2$.

In addition to transformation properties inherited from $\varphi$, a defining feature  of a \emph{paramodular form} is the exchange symmetry:
\be\label{eq:s11}
{\Phi}_k(\rho,\tau,z)= {\Phi}_k(t^{-1}\, \tau, t {\rho},z)~.
\ee
These transformation properties are encoded as follows. The paramodular group $\Gamma_t$ of level $t$  is defined as \cite{MR2208781}
\be\label{eq:defg2}
\Gamma_t :=\left[\begin{array}{cccc} 
	\Z & t\Z &\Z&\Z\\
	\Z &\Z&\Z&t^{-1}\Z\\
	\Z& t\Z&\Z&\Z\\
	t\Z&t\Z&t\Z&\Z
\end{array}\right] \cap Sp(4,\QQ)~.
\ee
It has an extension
\be\label{eq:defgammat}
\Gamma_t^+ = \Gamma_t \cup \Gamma_t V_t\ , \qquad
V_t = \frac{1}{\sqrt{t}}
\left(\begin{array}{cccc} 
	0&t&0&0\\
	1&0&0&0\\
	0&0&0&1\\
	0&0&t&0
\end{array}\right)\ .
\ee
Given a matrix $\gamma \in \Gamma_t^+$,  which we decompose into $2\times 2$ matrices as
\be\label{eq:gamma0}
\gamma=\twobytwo{{\bf A}}{{\bf B}}{{\bf C}}{{\bf D}}~,
\ee
the action of $\gamma$ on $\Omega$ is given by
\be\label{Phitrafo}
\tilde \Omega:=\gamma(\Omega)= ({\bf A}\Omega+{\bf B})({\bf C}\Omega+{\bf D})^{-1}\ .
\ee 
A meromorphic paramodular form ${\Phi}_k(\Omega)$ of weight $k$ is a meromorphic function on the Siegel upper half plane that satisfies
\be\label{eq:tz}
{\Phi}_k( \tilde \Omega )=\det({\bf C}\Omega+{\bf D})^k {\Phi}_k(\Omega)~.
\ee 
The exponential lift of a Jacobi form of weight 0 and index $t$ given in theorem~\ref{mainthm} is a meromorphic paramodular form with a character induced by $v^{24A}_\eta \times v^{2B}_H$, which means its transformation (\ref{Phitrafo}) has an additional phase $v(\gamma)$.

Moreover \cite{MR1616929} also establishes the divisors of $\Phi_k$, that is its zeros and poles. Since $\Phi_k$ is covariant under $\Gamma^+_t$, it is clear that we can group them into orbits of $\Gamma^+_t$.
These orbits are called \emph{Humbert surfaces} and are denoted by $H_D(b)$. $H_D(b)$ has a simple representative 
\be\label{eq:rep-pole}
a\tau +b z + t\rho =0 \ , 
\ee
where $a,b \in \ZZ$, $b$ is defined$\mod 2t$, and the discriminant $D$ is given by $D=b^2-4ta>0$.
The surfaces can thus be written as 
\be\label{eq:hdb}
H_D(b) = \pi^+_t(\{\Omega \in \H_2: a\tau + bz + t\rho =0 \} )\ ,
\ee
with $\pi^+_t$ the set of images of $\Gamma^+_t$. A general pole in $H_D(b)$ is then of the form
\be\label{quadPole}
tf(z^2-\tau\rho)+tc\rho+bz+a\tau+e=0\ .
\ee
Each such Humbert surface comes with multiplicity (or degree) $m_{D,b}$, which gives the order of the zero or pole. The total divisor of the exponential
lift is given by the Humbert surfaces
\be\label{Hmult}
\sum_{D,b} m_{D,b} H_D(b)~,
\ee
where the multiplicities $m_{D,b}$ are
\be
m_{D,b}= \sum_{n>0} c(n^2a,nb)\ ,
\ee
and $c(n,l)$ are the Fourier coefficients of the underlying Jacobi form $\varphi$ $(\ref{eq:nhJf})$. The multiplicity of the divisors is thus fixed by the polar terms of $\varphi$.

In this article we are interested in exponential lifts built from a wJf. For these forms, our goal is to extract the Fourier coefficients of
\be \label{eq:1overPhik}
\frac1{\Phi_k(\Omega)} = \sum_{m,n,l} d(m,n,l) p^m q^n y^l\ .
\ee
Since \eqref{eq:1overPhik} is meromorphic, we need to be more specific about the domain in which we expand it.  We choose
\be\label{eq:Imranges}
\textrm{Im}(\rho) \gg \textrm{Im}(\tau) \gg \textrm{Im}(z) > 0~,
\ee
which implies we expand first in $p$ then in $q$ and finally in $y$. From \eqref{explift}, note that 
\eqref{eq:1overPhik} has only terms with $n\geq -A$ and $m \geq -C$. And because $c(nr,l)$ in \eqref{explift} vanishes if $l < -\sqrt{4rtn+t^2}$, for a given term $p^m q^n$ there will only be finitely many $y$ powers with negative $l$. However there are infinitely many terms $y^l$ with positive $l$.

\section{Wall crossing method for negative discriminant terms}\label{sec:wall}

\subsection{Wall crossing and poles}\label{sec:genwall}

In this section we will highlight the main results of  the  methodology to extract the Fourier  coefficients of negative discriminant terms of paramodular forms developed in \cite{Sen:2011mh,Belin:2018oza}.  This will serve as the core to characterize the paramodular forms according to the behavior of their negative discriminant coefficients.

The goal is to evaluate the contour integral  
\begin{align} \label{cdef}
d(m,n,l)=\oint_{p=0} {dp\over 2\pi i p} \oint_{q=0} {dq\over 2\pi i q} \oint_{y=0} {dy\over 2\pi i y} \, \frac{1}{\Phi_k(\Omega)} p^{-m} q^{-n} y^{-l}
= \int_C d\rho d\tau dz \frac{e^{-2\pi i{\rm Tr}(\Omega Q)}}{\Phi_k(\Omega)} \,
\end{align}
for a negative discriminant term, i.e. 
\be
4mn-l^2 \, < \, 0~,
\ee	
and we defined the charge matrix
\be
Q =
\twobytwo{n}{{l\over2}}{{l\over 2}}{m}~.
\ee
$\Phi_k$ is a paramodular form built out of an exponential lift \eqref{explift}.  Taking the reciprocal is mostly conventional: It is common to define the first few coefficients of $\varphi$ as positive and $\Phi_k(\Omega)$ is thought of as a cusp form. 

Since $1/\Phi_k$ is meromorphic, we need to be more precise about the contour $C$ that we introduce. We want it to enclose $p=q=y=0$. For this choice we can take the real parts to be restricted to
\be\label{eq:re01}
0 \leq \text{Re}(\tau)~, ~\text{Re}(\rho)~,~ \text{Re}(z) <1 \,,
\ee
while for the imaginary parts we choose constant values
\be\label{eq:imlarge}
\text{Im} (\rho)~, ~\text{Im} (\tau)~, ~\text{Im}( z) \gg 0~.
\ee
In this way we define the contour $C$.
This gives the standard Fourier expansion of the SMF but note that in the $\rho,\tau,z$ plane, $C$ is not a closed contour.

There are two important observations in setting up the evaluation of \eqref{cdef}:
\begin{enumerate}
	\item $d(m,n,l)$  depends on the precise choice of the contour \cite{Cheng:2007ch}.  We can deform our contour within a domain without changing (\ref{cdef}) as long as we do not cross any poles when doing so. We therefore want to tessellate $\H_2$ into chambers whose boundaries are given by the poles of $1/\Phi_k$. To do this, first note that by choosing the imaginary parts of $\rho,\tau, z$ to be large as in $(\ref{eq:imlarge})$, the only poles in (\ref{quadPole}) that can contribute have $f=0$. That is, the tessellation is  characterized by the linear equation
	\be\label{eq:line1}
	t\,c\, \rho +b\,z+a\,\tau+e =0~, \qquad  e,a,b,c \in \Z~.
	\ee
	These lines will define regions (chambers) where we can unambiguously evaluate \eqref{cdef}.
	Next we will use the fact that the equation for the imaginary part of the poles (\ref{eq:line1}) is homogeneous. The boundaries of the chambers can thus be plotted in terms of the variables $\text{Im} (z)/\text{Im} (\tau)$ and $\text{Im} (\rho)/\text{Im} (\tau)$.
	
	We now choose the region {\bf R} in which the contour in \eqref{cdef} lies as follows:  \textbf{R} is the chamber which contains the points \eqref{eq:Imranges}. Recall that this means that the contour $C$ corresponds to the expansion of $1/\Phi_k(\Omega)$ first in $p$, then in $q$, and finally in $y$. 
	The lower boundary of {\bf R} corresponds to the minimal values of the integers in \eqref{eq:line1} that encloses the upper half plane \eqref{eq:uhp}. For example, for $t=1$ and Humbert surface $H_1(1)$, the lines defining the boundary of {\bf R} are 
	\be\label{eq:boundariesRH11}
	z=0~, \quad  z= \rho~, \quad  z= \tau~.
	\ee
	See Figure~\ref{eq:tessellationH11} for a plot of {\bf R} in this case, together with additional chambers.
	\begin{figure}[H]
		\centering
		{
			\begin{tikzpicture}[scale=1]
			\begin{axis}[
			unit vector ratio*=1 1 1,
			axis x line=center,
			axis y line=center,xlabel = $\frac{\mathrm{Im}(z)}{\mathrm{Im}(\tau)}$,
			ylabel = {$\frac{\mathrm{Im}(\rho)}{\mathrm{Im}(\tau)}$},  yticklabels=\empty, xticklabels=\empty,
			samples=100, every axis x label/.style={
				at={(ticklabel* cs:1)},
				anchor=west, scale=.5
			},
			every axis y label/.style={
				at={(ticklabel* cs:1.0)},
				anchor=south, scale=.5
			}
			]
			\addplot [name path=E,dashed, scale=.5, domain=0:1.15, line width=0.05mm] {x*x};
			\addplot [name path=A, domain=0:1] {x};
			\addplot [name path=N,white, domain=0:1] {1.28};
			\addplot[samples=50, smooth,domain=0:6,black, name path=three] coordinates {(1,1)(1,1.3)};
			\addplot [name path=B,domain=0:1/2] {.5*x};
			\addplot [name path=G,black, domain=0:1/2] {1/2*x};
			\addplot [name path=H, domain=0:1/3] {1/3*x};
			\addplot [name path=M,black, domain=0:1/4] {1/4*x};
			\addplot [name path=D,black, domain=1/2:1] {3/2*x-1/2};
			\addplot [name path=I,black, domain=1/3:1/2] {5/6*x-1/6};
			\addplot [name path=K,black, domain=1/4:1/3] {7/12*x-1/12};
			\addplot [name path=L,black, domain=1/5:1/4] {9/20*x-1/20};
			\addplot [name path=F,black, domain=0:1/5] {1/5*x};
			\addplot[teal!50] fill between[of=A and N];
			\addplot[magenta!50] fill between[of=A and G];
			\addplot[orange!50] fill between[of=H and G];      
			\addplot[violet!50] fill between[of=H and M];      
			\addplot[brown!50] fill between[of=F and M];                   
			\node[scale=1.3]  at (axis cs:  .3,  .7) {${\color{darkgray}\textrm{R}}$};
			\end{axis}
			\end{tikzpicture}
		}
		\caption{Tessellation of the Siegel upper half plane by Humbert surfaces belonging to $H_1(1)$ for $t=1$. The chamber {\color{teal}R} contains $p=q=y=0$ and is bounded by $(\ref{eq:boundariesRH11})$.}
		\label{eq:tessellationH11}
	\end{figure}
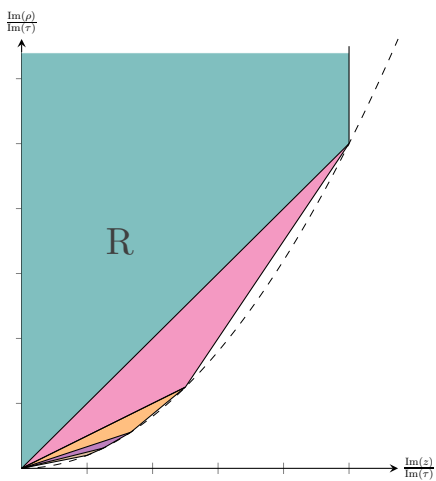

	\item A term that satisfies 
	\be\label{standardform}
	n < -A \qquad \textrm{or} \qquad m < -C\ , 
	\ee
	is defined to be of \emph{standard form}, and has $d(m,n,l)=0$ if expanded in {\bf R}.
	The crucial observation in \cite{Sen:2011mh} is that  a negative discriminant term, can always be brought to standard form by a suitable element of $\gamma\in\Gamma_t^+$. 
	
	More precisely, we can pick a transformation
	\be\label{eq:aa11}
	\tilde\gamma :=\twobytwo{{\bf A_\gamma}}{0}{0}{{\bf D_\gamma}}~, \quad {\bf A_\gamma}{\bf D_\gamma}^T=\mathds{1}_{2\times 2}~, \quad \det({\bf A_\gamma})=\pm1~ 
	\ee
	and given the restrictions in \eqref{eq:defg2}-\eqref{eq:defgammat}, we can parametrize the matrix ${\bf A_\gamma}$ as
	\be\label{eq:aa1}
	{\bf A_\gamma} = \twobytwo{a_1}{t b_1}{c_1}{d_1}~ , \quad a_1d_1-t b_1c_1=\pm1~,
	\ee
	with $a_1,b_1,c_1,d_1\in\ZZ$. Note that ${\bf A_\gamma} \in GL(2,\ZZ)$, and is allowed to have determinant $-$1.

	Acting with $ \tilde\gamma$ on our integration variable reads
	\be
	\tilde \Omega: =\tilde \gamma(\Omega)={\bf A_\gamma}\Omega {\bf D_\gamma}^{-1}\equiv \twobytwo{\tau_\gamma}{{z_\gamma}}{{z_\gamma}}{\rho_\gamma}\, .
	\ee
	As we will see, this leads to a new charge vector of the form
	\be\label{eq:gq1}
	\tilde Q := \tilde\gamma(Q) = {\bf D_\gamma}^{-1}Q {\bf A_\gamma} = \twobytwo{n_\gamma}{{l_\gamma\over2}}{{l_\gamma\over 2}}{m_\gamma}~.
	\ee
	In components this reads  as
	\bea\label{eq:tpoles1}
	\tau_\gamma&=&a_1^2 \tau +2t a_1b_1 z + t^2b_1^2\rho ~, \notag \\
	z_\gamma&=&a_1c_1 \tau + (a_1d_1+tb_1c_1) z + tb_1d_1 \rho~, \\
	\rho_\gamma&=&c_1^2 \tau + 2 c_1d_1 z + d_1^2 \rho \notag ~,
	\eea
	and
	\bea
m_{\gamma} &=& d_1^2m+t^2b_1^2  n+tb_1d_1 l~, \notag \\
n_{\gamma}&=&a_1^2 n+c_1^2 m+a_1 c_1 l~,  \\
l_{\gamma}&= &2 c_1 d_1 m+2 a_1 tb_1  n+(a_1d_1+tb_1c_1)l~ \notag ~.
	\eea	
By selecting ${\bf A_\gamma} $ appropriately, one can always set either $m_\gamma$ or $n_\gamma$ to be in standard form for negative discriminant terms. 
And it is important to observe that linear poles \eqref{eq:line1} get mapped to other linear poles under this set of transformations, as reflected in \eqref{eq:tpoles1}. Hence ${\bf A_\gamma} $ displaces us to different chambers. 
	
\end{enumerate}

We can use these two observations to compute the coefficient for any negative discriminant term in the following way:

\begin{enumerate}
	\item Given a charge vector $Q$, we find a transformation of the form \eqref{eq:aa11} such that $\tilde\gamma(Q)$ is of standard form.
	\item We map the image of the original contour $C$ to some contour
	\be
	\tilde C= \tilde\gamma( C) \ .
	\ee
	First note that $\tilde C$ lies in some different chamber \textbf{R'}. This is a result of the changes to the imaginary part of the variables. Secondly, we changed the integration range of the real parts. Note however that the total integration volume is unchanged: We have simply changed the cube (\ref{eq:re01}) into a parallelepiped of volume 
	$|\partial \tilde \Omega/\partial \Omega|=1$. 
	We can thus convert the real part of the integration range back into the original cube. This is the case because the whole integrand is invariant under shifts induced by ${\bf B}$ in \eqref{eq:gamma0}.
	
	\item Next we want to evaluate the contour integrals by closing them. To do this, we take $C$ to be the `top' and $\tilde C$ to be the bottom of a hypercube. We close this hypercube by adding `sides': that is, we keep the real parts fixed, and vary the imaginary parts such that they connect $\tilde C$ with the original contour $C$. We thus end up with a closed real 3-dimensional contour $K$. 
	\item We note that
	\be
	\int_K d\rho d\tau dz \frac{e^{-2\pi i{\rm Tr}(\Omega Q)}}{\Phi_k(\Omega)}
	=
	\int_C d\rho d\tau dz \frac{e^{-2\pi i{\rm Tr}(\Omega Q)}}{\Phi_k(\Omega)} \,.
	\ee
	To see this, note that only the `top' $C$ contributes. The contribution of the `bottom' $\tilde C$ vanishes because we can rewrite it as
	\begin{align}
	\int_{\tilde C} d\tilde\rho d\tilde\tau d\tilde z \frac{e^{-2\pi i{\rm Tr}(\tilde \Omega Q)}}{\Phi_k(\tilde\Omega)}
	&=
	\int_{C}d\rho d\tau dz \left|\frac{\partial \tilde \Omega}{\partial \Omega}\right|
	 \frac{e^{-2\pi i{\rm Tr}(\Omega \tilde Q)}}{\det(D)^k\Phi_k(\Omega)}\cr
	&= \pm \int_{C} d\rho d\tau dz \frac{e^{-2\pi i{\rm Tr}(\Omega \tilde Q)}}{\Phi_k(\Omega)} = 0\ ,
	\end{align}
	which vanishes because by construction
	$\tilde Q$ is in standard form, and $C$ is in \textbf{R}. Moreover the sides of $K$ mutually cancel each other out:  we always have two contributions where the real parts differ by 1. Consequently integers powers will cancel out because we integrate in opposite directions.
	\item
	Finally, the integral over the closed contour $K$ has the meromorphic integrand $1/\Phi_k(\Omega)$, so that it is given by the sum of all poles that we crossed in connecting $C$ with $\tilde C$.

\end{enumerate}

In summary,  the Fourier coefficient of a negative discriminant term is given by
\be \label{sumoverpoles}
d(m,n,l)= \sum_{{\bf p}_i} \frac{1}{2\pi i}{\rm Res} \left({q^{-n}p^{-m}y^{-l}\over \Phi_k}, {\bf p}_i\right) \,.
\ee
Here ``{\rm Res}'' stands for the residue integral around a pole ${\bf p}_i$, and we sum over a finite number of poles that we crossed.

An important point about the procedure described above is that it provides an algorithm to compute Fourier coefficients for $\Delta<0$ \textit{exactly}. This should be contrasted with the computation of (large and) positive discriminant term \cite{Sen:2007qy,Belin:2016knb} that involves various approximations that lead to asymptotic expressions for the coefficients $d(m,n,l)$. 

\subsection{Residue at ${\bf p}_i$}

Let us now discuss what poles can appear, and how to evaluate their residues. The possible linear poles are of the form \eqref{eq:line1}.
The order of the pole is dictated by $m_{D,b}$, and for simplicity we will take $m_{D,b}=1$ in what follows.  We will say a few words about higher order poles at the end of this section.

It is instructive to understand properties of the residues involved in \eqref{sumoverpoles}. For this purpose, let us evaluate the residue at the representative with $c=1$ in \eqref{eq:line1}, i.e.
\be\label{eq:p1}
{\bf p}_i :\quad p^t=y^{b}q^{-a} \,.
\ee
For $\textrm{Re}(\rho)\in[0,1)$, there are in fact $t$ such poles, corresponding to the different values of $e$, which we can take to be $e=0,...,t-1$. The values of $p$ then differ by $t$-th roots of unity $(\xi_t)^e$. To evaluate the residue, we pull out the factor $(1-p^t q^a y^{-b})^{-1}$. For each of the $e$ poles the residue of this factor is $\prod_{j=1}^{t-1}(1-\xi_t^j)^{-1}=t^{-1}$, and the contribution of the remaining factors where only powers of $p^t$ appear is (\ref{ResExp}) below. As there are $t$ poles, the total residue is thus (\ref{ResExp}). Therefore, by a slight abuse of notation, we will mean the residue at the pole ${\bf p}_i$ to mean the sum of the residues over all values of $e$.

We first evaluate the $\rho$ integral in \eqref{cdef} for  $m_{D,b}=1$; this gives
\be \label{res1pole}
{\rm Res} \left({q^{-n}p^{-m}y^{-l}\over \Phi_k}, {\bf p}_i\right) ={1\over 2\pi i}\int d\tau dz \, {\cal R}_i(\tau,z) q^{-n+ma/t} y^{-l-mb/t}~. 
\ee
The integrand here is given by
\begin{multline}\label{ResExp}
{\cal R}_i(\tau,z) = q^{-A+{a\over t}C} y^{-B+{b\over t}C} \prod_{l<0} (1-y^l)^{-c(0,l)}  \prod_{\substack{n>0\\ l\in\ZZ}} (1-q^n y^l)^{-c(0,l)} \prod_{\substack{r>0,n\geq0, l\in\ZZ\\ (n,l)\neq(ra,-rb)}}(1-q^{n-ar} y^{l+br})^{-c(nr,l)}\\
= (-1)^{2B}q^{-A+{a\over t}C} y^{B+{b\over t}C} \prod_{l>0} (1-y^l)^{-c(0,l)}  \prod_{\substack{n>0\\ l\in\ZZ}} (1-q^n y^l)^{-c(0,l)} \prod_{\substack{r>0,n\geq0, l\in\ZZ\\ (n,l)\neq(ra,-rb)}}(1-q^{n-ar} y^{l+br})^{-c(nr,l)}\\
=  (-1)^{2B}q^{-A+{a\over t}C} y^{B+{b\over t}C} \prod_{l>0} (1-y^l)^{-c(0,l)}   \prod_{\substack{n, L\in\ZZ\\(n,L)\neq(0,0)}}(1-q^{n} y^{L})^{-\fr(n,L)}\ ,
\end{multline}
where we have defined
\be\label{eq:deff}
\fr(n,L)\equiv \sum_{r=0}^{\infty} c(nr+ar^2, L-br)~.
\ee
Note that $t\rho = bz -a\tau$, which means that $b\text{Im} (z) - a\text{Im} (\tau)>0$. In particular this implies that we can expand (\ref{ResExp}) in $q^{-a}y^b$ and then $y$, which allows us to define the expansion of the residue as
\be\label{eq:rnl}
{\cal R}_i(\tau,z) = \sum_{n,l} r_i(n,l)q^n y^l\ .
\ee

We thus find that the contribution of the pole ${\bf p}_i$ to \eqref{sumoverpoles} is given by
\be\label{singleexact}
d_{{\bf p}_i}(m,n,l) = r_i(n-ma/t,l+mb/t)\ .
\ee
From (\ref{ResExp}) it is already clear that the growth of  $d_{{\bf p}_i}$ depends on the behavior of the $\fr(n,L)$. The asymptotic growth of \eqref{eq:deff} will lead to the distinction between slow growing and fast growing exponential lifts.

In Appendix \ref{app:high} we discuss how to proceed for higher order poles.  
The upshot is that we can still easily extract the leading large $m$ contributions to the residues. One may then wonder what the corrections that are subleading in $m$ look like, and what type of growth they exhibit. We discuss this question with a particular example in Appendix \ref{app:high} as well. In that case, we find that the growth is characterized by that of the divisor function $\sigma_1$. It would be interesting to understand the extent to which this holds for more general wJf. For the rest of the article however, we will continue to assume that we are dealing with simple poles.

\subsection{Example}\label{sec:example}

Let us illustrate this method via an example. Consider an exponential lift where the seed is the following wJf of index $t=2$,
\begin{align}
\phi_{0,2}(\tau,z)= \frac{1}{6}\left(\phi_{0,1}(\tau,z)^2 +5\phi_{-2,1}(\tau,z)E_4(\tau)\right)= \frac{1}{y^2} + 22+y^2 + \mathcal{O}(q)~.
\end{align}
From (\ref{Hmult}) we see that its exponential lift is a paramodular form of weight 11, i.e $\Phi_{11}$. The meromorphic form of interest is hence  $1/\Phi_{11}$, which
has single poles along the Humbert surfaces $H_1(1)$ and $H_4(2)$. 
We display parts of the tessellation of the Siegel upper half plane in figure~\ref{eq:tessellationH11H42} below.

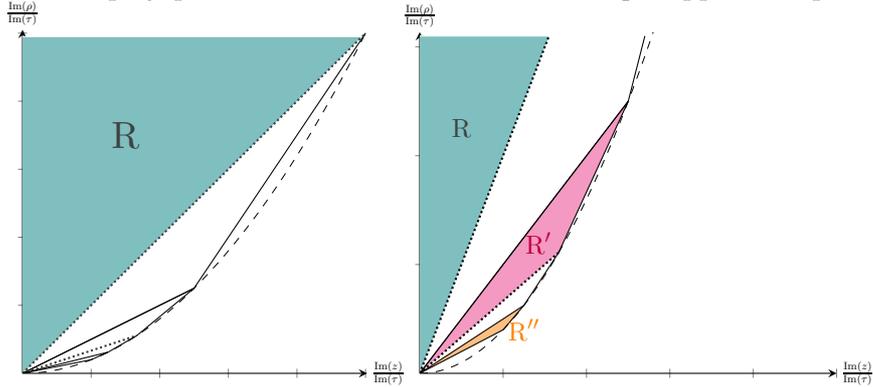
\begin{figure}[H]
\centering
{
\begin{tikzpicture}[scale=.8]
\begin{axis}[
unit vector ratio*=1 1 1,
axis x line=center,
  axis y line=center,xlabel = $\frac{\mathrm{Im}(z)}{\mathrm{Im}(\tau)}$,
    ylabel = {$\frac{\mathrm{Im}(\rho)}{\mathrm{Im}(\tau)}$},  yticklabels=\empty, xticklabels=\empty,
  samples=100, every axis x label/.style={
    at={(ticklabel* cs:1)},
    anchor=west, scale=.5
},
every axis y label/.style={
    at={(ticklabel* cs:1.0)},
    anchor=south, scale=.5
}
]
    \addplot [name path=E,dashed, scale=.5, domain=0:1.005, line width=0.05mm] {x*x};
     \addplot [name path=A,densely dotted, darkgray, line width=0.28mm, domain=0:1] {x};
      \addplot [name path=N,white, domain=0:1] {.99};
      \addplot [name path=B,domain=0:1/2] {.5*x};
        \addplot [name path=G,black, domain=0:1/2] {1/2*x};
         \addplot [name path=H, densely dotted, line width=0.28mm,darkgray,domain=0:1/3] {1/3*x};
         \addplot [name path=M,black, domain=0:1/4] {1/4*x};
       \addplot [name path=D,black, domain=1/2:1] {3/2*x-1/2};
       \addplot [name path=I,black, domain=1/3:1/2] {5/6*x-1/6};
        \addplot [name path=K,black, domain=1/4:1/3] {7/12*x-1/12};
         \addplot [name path=L,black, domain=1/5:1/4] {9/20*x-1/20};
        \addplot [name path=F,black, domain=0:1/5] {1/5*x};
           \addplot[teal!50] fill between[of=A and N];                   
          \node[scale=1.3]  at (axis cs:  .3,  .7) {${\color{darkgray}\textrm{R}}$};
    \end{axis}
    \end{tikzpicture}
}
{
{
}
}
{
\begin{tikzpicture}[scale=.8]
\begin{axis}[
unit vector ratio*=.38 1 .1,
axis x line=center,
  axis y line=center,
  xmin=0,xmax=1,
  xlabel =$\frac{\mathrm{Im}(z)}{\mathrm{Im}(\tau)}$,
    ylabel = {$\frac{\mathrm{Im}(\rho)}{\mathrm{Im}(\tau)}$},  yticklabels=\empty, xticklabels=\empty,
  samples=100, 
  every axis x label/.style={
    at={(ticklabel* cs:1.0)},
    anchor=west, scale=.5
},
every axis y label/.style={
    at={(ticklabel* cs:1.0)},
    anchor=south, scale=.5
}
]
    \addplot [name path=E,dashed, scale=.5, domain=0:.56, line width=0.05mm] {x*x};
     \addplot [name path=A,black, densely dotted, line width=0.28mm, domain=0:.31] {x};
      \addplot [name path=B,domain=0:1/2] {.5*x};
            \addplot [name path=N,white, domain=0:1] {.31};
       \addplot [name path=H,black, densely dotted, domain=0:1/3, line width=0.28mm] {1/3*x};
        \addplot [name path=G,black, domain=0:1/2] {1/2*x};
         \addplot [name path=M,black, domain=0:1/4] {1/4*x};
       \addplot [name path=D,black, domain=1/2:.54] {3/2*x-1/2};
        \addplot [name path=I,black, domain=1/3:1/2] {5/6*x-1/6};         
        \addplot [name path=K,black, domain=1/4:1/3] {7/12*x-1/12};
         \addplot [name path=L,black, domain=1/5:1/4] {9/20*x-1/20};        
        \addplot [name path=F,black, domain=0:1/5] {1/5*x};
        \addplot[magenta!50] fill between[of=G and H];
         \addplot[orange!50] fill between[of=F and M];
          \addplot[teal!50] fill between[of=A and N];
          \node[scale=.9]  at (axis cs:  .1,  .225) {${\color{darkgray}\textrm{R}}$};
       \node[scale=.9]  at (axis cs:  .285,  .12) {${\color{purple}\textrm{R}'}$};
       \node[scale=.9]  at (axis cs:  .25,  .04) {${\color{orange}\textrm{R}''}$};
    \end{axis}
    \end{tikzpicture}
}
\caption{{\it Left:} Tessellation of the Siegel upper half plane (dashed line) by some  of the Humbert surfaces being part of  $H_1(1)$ (solid line) and $H_4(2)$ (dotted). The turquoise  chamber {\color{teal}R} contains $p=q=y=0$. 
{\it Right:} This figure zooms into the origin of the left figure, displaying the chambers we are moving by applying the transformation $A_{\gamma_1}$ ({\color{magenta}$\textrm{R}'$ magenta}) and $A_{\gamma_2}$ ({\color{orange}$\textrm{R}''$ orange}).}
\label{eq:tessellationH11H42}
\end{figure}
We will evaluate two different Fourier coefficients $1/\Phi_{11}$. To start, we first consider  the coefficient $d(4,-1,-3)$, which is a non-trivial term with negative discriminant in $1/\Phi_{11}$.
Following \eqref{eq:aa11}-\eqref{eq:aa1}, the element 
\begin{align}\label{eq:gamma1}
A_{\gamma_1}=    \begin{pmatrix} 
1 & 2 \\
0 & 1 \\
\end{pmatrix}
\end{align}
brings it into standard form $(\ref{standardform})$. This element deforms the contour ${\cal C}$ in $\bf{R}$ such that it crossess the poles $2\rho=2z$ --belonging to $H_4(2)$-- and $4\rho=2z$, --belonging to $H_1(1)$-- ending in the magenta chamber $\bf{R'}$ in Fig.\,\ref{eq:tessellationH11H42}. The residue at $4\rho = 2z$ is zero and we simply obtain from \eqref{sumoverpoles} the expression 
\be\label{eq:residuec4m1m3}
d(4,-1,-3)= \frac{1}{2\pi i}{\rm Res} \left({q^{1}p^{-4}y^{3}\over \Phi_{0,2}}~, \bf{p^2=y^2}\right)~.
\ee
This residue is of the form \eqref{res1pole}, and by expanding  \eqref{ResExp} in $q$ and then $y$ we read off 
\begin{align}
r_{2\rho=2z}(-1,1) =-24~,
\end{align}
where $r_i$ is defined in \eqref{eq:rnl}. Hence $d(4,-1,-3)=-24$.
It is straightforward to check that this agrees with the value obtained by a direct expansion of $1/\Phi_{11}$.

Next we consider the coefficient $d(4,-1,-1)$. It has again negative discriminant, but it is not brought into standard form by $(\ref{eq:gamma1})$. Instead we need the transformation 
\begin{align}
A_{\gamma_2}=    \begin{pmatrix} 
1 & 4 \\
0 & 1 \\
\end{pmatrix}~.
\end{align}
We now not only cross the poles $2\rho=2z$ and $4\rho=2z$ as before, but also  the two poles $6\rho=2z$ --in $H_4(2)$-- and $8\rho=2z$ --in $H_1(1)$-- such that we land in the orange chamber $\bf{R''}$ in Fig.\,\ref{eq:tessellationH11H42}. The residues along these two latter poles vanish. However, this time the residue of the pole $4\rho=2z$ no longer vanishes. This means $d(4,-1,-1)$ is given by the sum of two residues
\begin{align}\label{eq:doublepole}
d(4,-1,-1)=\frac{1}{2\pi i}\left({\rm Res} \left({q^{-4}p^{1}y^{1}\over \Phi_{0,2}}~, \bf{p^2=y^2}\right)+ {\rm Res} \left({q^{-4}p^{1}y^{1}\over \Phi_{0,2}}~, \bf{p^4=y^2}\right) \right)~.
\end{align}
From $(\ref{eq:doublepole})$ it follows
\begin{align}
d(4,-1,-1)&= r_{2\rho=2z}(-1,3)+ r_{4\rho=2z}(-1,1)= -324+1 = -323\ .
\end{align}
This again agrees with the direct expansion of $1/\Phi_{11}$.

Let us try to summarize the findings from this example:
\begin{enumerate}
	\item To obtain the coefficients, in general we cross multiple poles, coming from Humbert surfaces involved in the exponential. However,  the residues of some of the poles tend to vanish, which eases the procedure.
	\item For coefficients with very negative discriminant, the full coefficient comes from the residue of a single pole. That is, we should be able to find a range of values for $m,n,l$ such that $d(m,n,l)$ is in the \emph{single pole regime}, \ie gets contribution from only a single pole. We will work this out in the next section.
\end{enumerate}

\subsection{Single pole regime}\label{sec:single}

Our aim in this section is to identify negative discriminant terms where the residue of only one pole contributes to $d(m,n,l)$ in \eqref{sumoverpoles}. 
In view of what follows, let us be slightly more precise about the nature of these terms. Let us make the assumption that the most polar term of the seed is 
\be\label{eq:seed22}
q^0 y^{-b_0}~,
\ee
and comes with a positive coefficient. For a general $\varphi$, the most polar term could of course also be of the form $q^{a_0}y^{-b_0}$. We leave the analysis of this slightly more general case for future work.

Neglecting for a moment the terms in the product formula \rref{explift} that have $m=0$, for large but fixed $m$ and $n=0$ the term with the most negative power of $y$ is $q^0 y^{-mb_0/t}$. At the same time this term also has minimal discriminant for this value of $m$. Terms nearby  this minimal discriminant are of the form
\be\label{coeffsinglepole}
d(m,n,-mb_0/t +\ell)~, 
\ee
where 
\be\label{coeffsinglepole1}
n\sim O(m^0) ~, \quad\ell \sim O(m^0) ~, \quad m\gg1~.
\ee
In other words, we are interested in keeping $n$ and $\ell$ fixed while $m$ is parametrically large. We will show that  the algorithm to obtain $d(m,n,l)$ described in Sec.\,\ref{sec:genwall} simplifies significantly in the regime \eqref{coeffsinglepole1}.

The algorithm simplifies in the following sense. Given our assumption about the most polar term of the seed, i.e. \eqref{eq:seed22},  there is a simple pole $\mathbf{p}_0$ with
\be
b_0 z + t\rho =0 \ , 
\ee
which belongs to the Humbert surface with maximal discriminant (as usual we include the other poles with $b=b_0, a=0, e\neq 0$ here). We then claim the following:
For coefficients satisfying \eqref{coeffsinglepole}--\eqref{coeffsinglepole1},  $d(m,n,l)$ only receives a contribution from the pole $\mathbf{p}_0$. That is
\be\label{dsinglepole}
d(m,n,l) = r_0(n,l+mb_0/t)~, \quad l=-mb_0/t +\ell~,
\ee
where $r_0$ comes from the expansion of
\be\label{singlepoleresidue}
{\cal R}_0(\tau,z) =  (-1)^{2B}q^{-A} y^{B+{b\over t}C} \prod_{l>0} (1-y^l)^{-c(0,l)}   \prod_{\substack{n\geq 0, L\in\ZZ\\(n,L)\neq(0,0)}}(1-q^{n} y^{L})^{-\fr(n,L)}\ ,
\ee
and $\fr(n,L)$ is given by \eqref{eq:deff} with $a=0$. 
The coefficients in $1/\Phi_k$ satisfying \eqref{dsinglepole} will be denoted as coefficients in the single pole regime.  

To establish \eqref{dsinglepole}, we need to prove that no other pole $\mathbf{p}_i$ of the linear form (\ref{eq:line1}) can make a contribution.
Let us first investigate the case of $a=0,c=1,e=0, b< b_0$, so that the 
pole is given by $p^t y^{-b}$. We can then rewrite the residue as
\begin{multline}\label{ResExpa0}
{\cal R}_i(\tau,z) =  (-1)^{2B}q^{-A} y^{B+{b\over t}C} \prod_{l>0} (1-y^l)^{-c(0,l)}   \prod_{\substack{n\geq 0, L\in\ZZ\\(n,L)\neq(0,0)}}(1-q^{n} y^{L})^{-\fr(n,L)}\ , \\
 =  (-1)^{2B}q^{-A} y^{B+{b\over t}C} \prod_{l>0} (1-y^l)^{-c(0,l)}   \prod_{\substack{n> 0, L\in\ZZ\\(n,L)\neq(0,0)}}(1-q^{n} y^{L})^{-\fr(n,L)}\ \prod_{\substack{L\in\ZZ-\{0\}}}(1- y^{L})^{-\fr(0,L)} ,
\end{multline}
By the remarks above, we first expand ${\cal R}_i$ in positive powers of $q$, and then in positive powers of $y$. Note that the last product can contain a finite number of factors with negative powers of $y$. These we will also expand in positive powers of $y$, possibly leading to an additional simple negative power pre-factor of $y$. The crucial observation is that negative powers of $y$ can only appear in conjunction with positive powers of $q$.

To obtain the contribution to $d(m,n,l)$ we use  
\be\label{bRes}
d_{{\bf p}_i}(m,n,l) = r_i(n,l+mb/t)\ .
\ee
This means that we are looking for the term $q^N y^L$ in the expansion of  (\ref{ResExpa0}) with
\be
L = \ell -m(b_0-b)/t\ , \qquad N = n\ .
\ee
For large $m$ this means that $L$ is negative and parametrically large, but $n$ is not parametrically large. We now want to argue that there is no such term.

To see this we note that the product (\ref{ResExpa0}) may have a negative power of $y$ coming from the pre-factor. Other than that, the only negative powers that appear come from the factors with exponent $\fr(n,L)$ and $n>0$. On the other hand, the definition \eqref{eq:deff} combined with the polarity constraint ($c(n,l)=0$ if $4tn-l^2 < -b^2$) implies that
\be\label{Lsinglepolebound}
\fr(n,L) = 0 \qquad \textrm{if}\ L < -\frac{b^3}{4 n t}-\frac{n t}{b}~.
\ee
This means that a term $q^n y^{L}$ can only have negative and parametrically large $L$ if $n$ is also parametrically large.
This establishes that no term from such a pole will make a contribution to $d(m,n,l)$ in the regime \eqref{coeffsinglepole1}.\footnote{ It is also clear that this argument does not change if we allow for $e\neq0$: In that case, the factors will at most pick up a root of unity, which does not change the exponents.}

Next consider the contribution of a pole $\mathbf{p}_i$ with $c=1,a>0,b<b_0,e=0$, \ie
the  contribution of the pole ${\bf p}_i$ to the coefficient is given by
\be\label{singleexact}
d_{{\bf p}_i}(m,n,l) = r_i(n-ma/t,l+mb/t)\ ,
\ee
and the residue itself is given by
\begin{multline}\label{ResExpa}
{\cal R}_i(\tau,z) 
= (-1)^{2B}q^{-A+{a\over t}C} y^{B+{b\over t}C} \prod_{l>0} (1-y^l)^{-c(0,l)}  \prod_{\substack{n>0\\ l\in\ZZ}} (1-q^n y^l)^{-c(0,l)} \prod_{\substack{r>0\\n\geq0\\ l\in\ZZ}}(1-q^{n-ar} y^{l+br})^{-c(nr,l)}~.
\end{multline}
We are then looking for terms $q^N y^L$
\be
L = \ell -m(b_0-b)/t\ , \qquad N = n-ma/t\ .
\ee
This means that we are extracting a parametrically large negative coefficient of $q$ in (\ref{ResExpa}). This coefficient will have to come from the expansion in terms of $y^b q^{-a}$, which means that the contribution will have a positive power of $y$ which is parametrically large, namely of the form $y^{bm/t} q^{-ma/t}$. In order to get a parametrically large negative power of $y$, we would need a contribution of the form $y^{-mb_0/t}$ without a parametrically large power of $q$. From the arguments above we know this is not possible. The same argument still holds for $e\neq 0$.

Finally, for any pole with $c>1$ we can repeat the computation of the residue by simply dividing (\ref{eq:line1}) by $c$. This will lead to a new $b$ which will be smaller than $b_0$, which in turn cannot make a contribution in the regime \eqref{coeffsinglepole1}. This exhaust all possible poles that could contribute to $d(m,n,l)$, and hence we have shown that in the regime \eqref{coeffsinglepole1}, the claim \eqref{dsinglepole} is true.


\section{Fast and slow growing SMFs}\label{s:Bad}

\subsection{Overview}
In the previous section we established the following: In the single pole regime \eqref{coeffsinglepole1}, the coefficients $d(m,n,l)$ are given by the residue (\ref{dsinglepole}). In particular, their growth is determined by the exponent $\fr(n,l)$ \eqref{eq:deff} with $a=0$. In this section we will determine the behavior of $\fr(n,l)$.

As we will see, it can have only two types of asymptotic behavior:
\begin{enumerate}
\item $\fr(n,l)$ is bounded as  a function of $n$ and $l$. More precisely, it only  takes a finite number of different values. The residue is thus essentially a ratio of $\theta$-functions, and the $r_i(n,l)$ in \eqref{eq:rnl} grow accordingly. We will call such wJf and their exponentially lifted SMF \emph{slow growing}.
\item $\fr(n,l)$ is unbounded and grows exponentially with  $n$ and $l$. In that case its growth is roughly of the form
\be\label{eq:fr333}
\fr(n,l)\sim \exp 2\pi \sqrt{4\alpha (tn^2/b^2+nl/b)}
\ee
for some $\alpha \leq 1$. 
We will call such wJf and their exponentially lifted SMF \emph{fast growing}.
\end{enumerate}
Note that we distinguish the two cases by the growth behavior of the $\fr$, rather than the $d(n,m,l)$. 
For the slow growing case, it is straightforward to extract the growth of the $d(m,n,l)$ from the $\fr(n,l)$, since the generating function is a ratio of theta-like functions. For the fast-growing case, it is in principle more subtle. We expect however that the growth of the $d(m,n,l)$ is dictated by the growth of the exponents themselves. This is known to be the case for the bosonic partition function \cite{Keller:2011xi}, and we do not expect any changes here. It would interesting to investigate this carefully.
Under this assumption the growth of the $d(m,n,l)$ can then be inferred from this expression through \rref{singleexact} and \eqref{eq:fr333}. Namely, in the first case they will grow roughly as $\exp \sqrt{n}$, whereas in the second case they will grow as $\exp n$.

\subsection{Generic behavior of $f(n,l)$}

Instead of working with the $\fr(n,L)$ in \eqref{eq:deff} it is more convenient to introduce
\be \label{fsum}
f(n,l) = \sum_{r\in\Z} c(rn,l-br) \ ,
\ee
which, for $a=0$, is related to $\fr$ by
\be
f(n,L) = \fr(n,L) + \delta_{n,0}\sum_{r>0}c(0,L+br)\ .
\ee
As we infer properties of $f(n,L)$, it will be simple to translate them to $\fr(n,L)$, since they differ by a finite (constant) number of terms.

Let us first discuss the `generic' behavior of the $f(n,L)$.
Here `generic' means that we assume no large cancellations between the coefficients $c(n,l)$ in the sum $(\ref{fsum})$. That is, the sum is well approximated by its largest term. To find the largest term, we use the fact that for large discriminant, the asymptotic behavior of the coefficients is given  by  
\be\label{eq:cardy}
c(n,l)\sim \exp  \pi\sqrt{{|\Delta_{\rm min}|\over t^2}(4tn-l^2)} ~ ,
\ee 
where $\Delta_{\rm min}$ is the maximal polarity of the seed wJf; see for instance appendix B of \cite{Belin:2016knb} for a derivation of \eqref{eq:cardy}. Approximating $c(n,l)$ by \eqref{eq:cardy}, we find that  the maximal term in (\ref{fsum}) occurs for $r=(2tn+bL)/b^2$ and gives
\be\label{fR}
c(n,L)\sim \exp 2\pi \sqrt{{|\Delta_{\rm min}|\over b^2 \,t}( n^2 t + b nL) }~. 
\ee
A `generic' $f(n,L)$ then grow as (\ref{fR}).
Assuming that the largest discriminant comes from our term $q^0y^{-b_0}$, we have $\Delta_{\rm min}=-b_0^2$. For terms with $L=0$ we then simply have
\be
f(n,0) \sim \exp 2\pi n\ ,
\ee
that is Hagedorn growth with slope $2\pi$. This is in accordance with the bounds found in \cite{Keller:2011xi}. 

As we will see in the next section, there are indeed many instances with no cancellations, and the $f(n,l)$ grow as in (\ref{fR}). However, there are interesting cases where the growth can be much slower.

\subsection{Generating functions}

Let us now study the growth of the $f(n,l)$ in more detail. Given a wJf $\varphi$ of weight 0 and index $t$ whose coefficients are given by $c(n,l)$, we need to  evaluate \eqref{fsum}.
 Our strategy will be  to build generating functions for the $f(n,l)$. These turn out to be modular functions of certain congruence subgroups, which allows us to extract the behavior of their coefficients.

First let us define
\be\label{Deltab}
M = tn^2/b^2+nl/b\ .
\ee
In general, $M$ is not an integer. We therefore define
\be
n_b := n \mod b \in \{0,1,\ldots,b-1\}~,
\ee
and using the identity
\be\label{coeffshift}
c(n,r) = c(n+r\lambda+t\lambda^2, r+2t\lambda) \ , \qquad \lambda \in \Z\ ,
\ee
and picking $\lambda = (n-n_b)/b$, we obtain
\be
c(rn,l-br)= c(t(n-n_b)^2/b^2+l(n-n_b)/b+n_b r, l-br+2(n-n_b)t/b)~,
\ee
where clearly the first argument is an integer. Next we replace the summation over $r$ by a summation over $\hat m$ in \eqref{fsum}, where 
\be
-\hat m := br-2(n-n_b)t/b-l = br -k \ .
\ee
The sum is now no longer over integers, but over $b\Z+k$ with $k=2(n-n_b)t/b+l$. In terms of the new variables $(M, n_b, k)$, the function in \eqref{fsum} reads
\be
f(M,n_b,k)=\sum_{\hat m \in b\Z+k}c(M -n_b \hat m/b-n_b^2t/b^2, \hat m)
\ee
Note that for fixed $n_b$, $M \in \Z +n_bk/b+ n_b^2t/b^2$. We can define the generating function
\be
F_{n_b,k}(\tau):=\sum_{M}f(M,n_b,k)q^M = \sum_{\substack{M\\ \hat m \in b\Z+k}}c(M -n_b \hat m/b-n_b^2t/b^2, \hat m)q^M\ ,
\ee
which we rewrite as
\be
F_{n_b,k}(\tau)=q^{n_b^2t/b^2}\sum_{\substack{\tilde \Delta\in\Z\\\hat m \in b\Z+k}} c(\tilde M,\hat m) q^{\tilde M}q^{n_b\hat m/b}
=q^{n_b^2t/b^2}\sum_{\substack{\tilde M\in\Z\\ \hat m\in\Z}} c(\tilde M,\hat m) q^{\tilde M}q^{n_b\hat m/b}\delta^{(b)}_{\hat m,k}~.
\ee
Here $\tilde M$ is the integral part of $M$, and  $\delta^{(b)}_{\hat m,k}$ is the $b$-periodic Kronecker delta. Using 
\be
\delta^{(b)}_{\hat m, k}={1\over b}\sum_{j=0}^{b-1} e^{2\pi i(\hat m - k) j/b} ~,
\ee
we can finally rewrite this as
\be\label{Fexpr}
F_{n_b,k}(\tau) = \frac1b\sum_{j=0}^{b-1}q^{n_b^2t/b^2} \varphi_{0,t}\left(\tau, \frac{n_b\tau + j}{b}\right) e^{-2\pi i kj/b }
= \frac1b\sum_{j=0}^{b-1}\chi_{n_b,j}(\tau) e^{-2\pi i kj/b }\ ,
\ee
where  $\varphi_{0,t}$ is the seed in the exponential lift, and as specializations of a two variable wJf $\varphi$, we have defined the functions
\be \label{eq:chi19}
\chi_{r,s}(\tau):=q^{tr^2/b^2}\varphi(\tau, (r\tau+s)/b)\ , \qquad r, s= 0,1,\ldots b-1~.
\ee
Note that we have $\chi_{r,s}(\tau)=\chi_{r,s+b}(\tau)=\chi_{r+b,s}(\tau)e(st/b)$, which is why we restricted the indices to the range above.
To quantify the growth of $f(M,n_b,k)$, we thus need to understand the properties of $\chi_{r,s}(\tau)$. As we will establish in the next section, they turn out to have good modular transformation properties, which allows us to extract the growth of their coefficients. We will turn to this now.

\subsection{Specialized transformation properties}
In what follows, we need to understand modular transformation properties of specialized versions of wJf. To this end we use the properties of wJf described in appendix~\ref{app:wjf}.

Given a wJf $\varphi$ of weight 0 and index $t$, for a fixed $b \in \NN$, we define the specialized forms \eqref{eq:chi19}.
Under the $S$ element of $SL(2,\Z)$, we have
\begin{align}
\chi_{r,s}(S\tau)&:= \chi_{r,s}(-1/\tau)\cr
&= e(-tr^2/b^2\tau)\varphi\left(-\frac1{\tau}, \frac{(s\tau - r)/b }{\tau} \right)
= e(-2srt/b^2)q^{ts^2/b^2}\varphi(\tau,(s\tau-r)/b)) \\ \nonumber
&= e(-2srt/b^2)\chi_{s,-r}(\tau)
\end{align}
and the $T$ transformation gives 
\begin{align}
\chi_{r,s}(T\tau)&:= \chi_{r,s}(\tau+1)
\cr &=e(tr^2/b^2)\chi_{r,s+r}(\tau)\ .
\end{align}
Note that we indeed have
\be
\chi_{r,s}(S^2\tau)=\chi_{-r,-s}(\tau)=\chi_{r,s}(\tau)~,
\ee
and
\be
\chi_{r,s}((ST)^3\tau)= \chi_{-r,-s}(\tau)=\chi_{r,s}(\tau)~,
\ee
which establishes that the $\chi_{r,s}(\tau)$ form a (in general reducible) representation of $SL(2,\Z)$.
By using a sequence of such transformations, \ie by transforming by  a suitable
element of $SL(2,\Z)$ we can relate any $\chi_{r,s}(\tau)$ to some $\chi_{r',0}(\tau)$: Simply use $T$ until $s<r$, and then use $S$ to exchange $s$ and $t$. Continue until you reach $s=0$.

The $\chi_{r,s}(\tau)$ are of course holomorphic on $\H$. Their only possible  singularities are at the cusps $\{i\infty, \mathbb{Q} \}$.
\begin{defn}
We call a wJf $\varphi$ to be of \emph{slow growth type} if 
\be\label{sugracondition}
\forall ~ r, s = 0,1,\ldots b-1\qquad \chi_{r,s}(\tau) \quad \textrm{is regular for}\ \tau\to i\infty ~.
\ee
\end{defn}
Our claim is that if $\varphi$ is of slow growth type, then all $\chi_{r,s}(\tau)$ are constant.
To see this, first note that by theorem 1.3 in \cite{MR781735}, $\chi_{r,s}(\tau)$ is invariant under some congruence subgroup $\Gamma_{r,s} \subset SL(2,\Z)$.
$\chi_{r,s}(\tau)$ is holomorphic away from the cusps. Any cusp can be mapped to $i\infty$ by a suitable $SL(2,\Z)$ transformation, under which $\chi_{r,s}(\tau)$ transforms into some $\chi_{r',s'}(\tau)$, which by (\ref{sugracondition}) however is regular. $\chi_{r,s}(\tau)$ is thus holomorphic on the compactification of $\H/\Gamma_{r,s}$, which in particular implies that it is constant. 
From (\ref{Fexpr}) it follows that all $F_{n_b,k}(\tau)$ are constant. 
This in turn implies that the $f(n,l)$ vanish unless $M=0$. It follows that the $f(n,l)$ are indeed bounded, so that we are in the slow growing case.

For the slow growing case, we can in fact give closed form expressions for the $f(n,l)$: 
They can be computed as described above and vanish unless $M=0$, that is
\be
n=0 \qquad \textrm{or}\qquad tn+bl=0\ .
\ee
Moreover they only depend on $n_b := n \mod b$ and on $k_b := k \mod b$. 
 In total we have
\begin{align}
f(n,l)= \left\{ \begin{array}{ccl} 
	\sum\limits_{\hat m \in b\Z-l-n_bt/b}c(-n_b \hat m/b-n_b^2t/b^2, \hat m) & : &tn+bl=0
	\ \textrm{or}\  n=0
	\\ 
	0 & :& \textrm{else}
\end{array}\right.
\end{align}
Note that we automatically have $n_bt/b\in \Z$. Also, since $c(-n_b \hat m/b-n_b^2t/b^2, \hat m)$ has negative discriminant for all $\hat m$, there are at most $b^2$ constants that enter $f(n,l)$. From this it follows that the generating function for the $d(m,n,l)$ in the single pole regime, i.e. \eqref{singlepoleresidue}, is essentially a ratio of $\theta$-type functions, which implies that the $d(m,n,l)$ indeed exhibit  slow growth behaviour.

Conversely, assume that $\chi_{r,s}(\tau)$ has a pole at $i\infty$, 
\be
\chi_{r,s}(\tau) \sim q^{-\alpha}\ ,
\ee
then by applying an $S$ transformation (i.e. the usual Cardy argument) we obtain an exponential growth for the coefficients of $\chi_{s,-r}(\tau)$, namely
\be\label{alphaCardy}
\sim \exp 2\pi \sqrt{4\alpha \tilde M} \sim \exp 2\pi \sqrt{4\alpha (tn^2/b^2+nl/b)}\ .
\ee
As we can see, this implies that we are in case 2 described above.

\section{Examples}\label{sec:example-slow-fast}
\subsection{How to identify slow growing SMF}

Let us now discuss how to determine in practice if a given $\varphi$ leads to slow growth or to fast growth, and work out some examples. From the above, this means we need to check the behavior of $\tau\to 0$ for the functions $\chi_{n_b,j}(\tau)$, which is equivalent to working out the behavior of $\tau\to i\infty$ for the function $\chi_{j,n_b}(\tau)$. To do this, note that a term of the form $q^ny^{-l}$ in $\varphi$ leads to a term $q^\beta$ for $\chi_{j,n_b}$, where
\be\label{eq:betaterm}
\beta = tj^2/b^2 +n -l j/b = \frac{t}{b^2}j(j-lb/t)+n\ .
\ee
Our interest in quantifying the conditions on $c(n,l)$ such that we encounter $\beta<0$. For this it is useful to consider the maximisation of $-\beta$: 
\begin{align}\label{alphacriterion}
\alpha &:= \max_{j=0,\ldots,b-1} [-\beta]\cr 
&=\max_{j=0,\ldots,b-1}
\left[-t\left(\frac{j}{b}-\frac{l}{2t}\right)^2-{1\over 4t}\le(4tn-l^2\ri) \right]\ .
\end{align}
If and only if $\alpha>0$, the term $q^n y^{-l}$ leads to a singular behaviour of the form $q^{-\alpha}$ in $\chi_{j,n_b}$.  In this case there is an exponential growth in the coefficients of $\chi_{n_b,j}$ as in \eqref{alphaCardy}. If $\alpha$ is non-positive for all polar terms, then $\chi_{j,n_b}$ is a constant and it leads to the slow growth case. 
Note that for a non-polar term $q^n y^{-l}$, we automatically get a non-positive $\alpha$. 
To check if a form leads to fast growth or slow growth, we therefore need to test (\ref{alphacriterion}) only for the polar terms of $\varphi$.
Moreover from (\ref{eq:betaterm}) it is clear that $\beta \geq 0$ if $l\leq 0$, and from (\ref{alphacriterion}) that $\beta\geq 0$ if $l\geq 2t$. In fact, if $t<l<2t$, then there is a corresponding term with $\tilde l = l-2t$ and $\tilde n$ which has the same polarity, which by $c(\tilde n, \tilde l) = c(\tilde n, -\tilde l)$  also has a corresponding term with $0<\tilde l< t$. This term then provides the same value of $\alpha$ because of
\be
\left(\frac{j}{b}-\frac{l}{2t}\right)^2
=\left(\frac{b-j}{b}-\frac{\tilde l}{2t}\right)^2\ .
\ee
The upshot is thus that we only need to check (\ref{alphacriterion}) for polar terms $q^n y^{-l}$ with $0<l \leq t$, of which there are only finitely many.

For a given form $\varphi$, this is of course a simple procedure. Before working it out for several examples with small $t$, let us discuss some more general properties. 
Consider the term $y^{-b}q^0$.
In principle, $\beta$ is maximized for $j=b^2/2t$, giving $\alpha=b^2/4t$, which
gives exponential growth
\be
\exp 2\pi \sqrt{4\alpha (tn^2/b^2+nl/b)} \,.
\ee
We see that this is exactly the non-cancellation case that led to (\ref{fR}).
The reason why we sometimes have either a different slope, or slower growth,  is that this optimal value of $j$ cannot be attained:
This value of $j$ can only be attained if 
\be
\frac{b^2}{2t} \in \Z\ .
\ee
If this condition is not satisfied, then we will get an exponential growth with a smaller coefficient, or even slow growth. 

Let us now discuss the circumstances that distinguish slow from fast cases. Inspecting \rref{eq:betaterm}, we see that a {\it sufficient condition for fast growth} is that the most polar term in $\varphi$ satisfies
\be\label{eq:bad}
b^2 > t\ .
\ee
Then the maximal polarity term $q^{0}y^{-b}$ will give $\alpha>0$, and hence leads to fast growth.

From this, we can just as well derive a {\it necessary condition for slow growth:}
\be\label{eq:nc1}
b^2 \leq t\ .
\ee
This condition indicates that the term $q^{0}y^{-b}$ does not lead to fast growth, i.e. we will have $\alpha\leq0$. Moreover, any term in $\varphi$ of the form $q^0y^{-l}$ will lead to $\beta>0$ in \eqref{eq:betaterm}. Therefore, given \eqref{eq:nc1}, to find all slow growing forms for a given index $t$, we need to:
\begin{itemize}
\item For each $b\leq \sqrt{t}$, construct all wJf $\varphi$ with terms of polarity $\Delta\geq -b^2$.
\item For those $\varphi$, check that negative discriminant terms in $\varphi$ of the form $q^{n}y^{-l}$, with $n\neq0$, do not lead to a term in $\chi_{j,n_b}$  of the form $q^{-\alpha}$ with $\alpha>0$.
\end{itemize}

\subsection{Examples with small index $t$}
Let us discuss some examples of low index. For $t\leq 4$, all polar terms are of the form $q^0 y^{-b}$, $b=1,\ldots t$. Moreover for these values of $t$ there is exactly one weak Jacobi form whose only polar term is given by $q^0 y^{-b}$. In table~\ref{t:slope} we have worked out the growth behavior for all such forms. \begin{table}
	\centering
	\begin{tabular}{|cccc|}
		\hline
		$t$ &  $b$ & $j$ & $\alpha$ \\
		\hline
		1 &1 &0 & 0\\
		\hline
		2 & 2 & 1 & $1/2$\\
		\hline
		3 & 1 & 0 &0 \\
		3 &2 & 1 & 1/4\\
		3 &3 & 1 & 2/3\\
		\hline
		4 & 1 & 0 &0  \\
		4 & 2 & 1 & 0\\
		4 & 3 & 1 &  5/9\\
		4 & 4& 2 & 1\\
		\hline
	\end{tabular}
	\caption{Slope for various examples up to $t=4$. $j$ is the value that maximizes $\beta$. A slope of 0 indicates slow growth. \label{t:slope}}
\end{table}

For $t\geq5$, the situation becomes more complicated. On the one hand, there are now also polar terms $q^n y^{-l}$ with $n>0$. 
From \eqref{eq:nc1} we know that necessarily $b\leq\sqrt{t}$, since otherwise we will automatically have fast growth. 
On the other hand, it is critical to keep track of the dimension of the space of wJf and their polar terms. Following \cite{Gaberdiel:2008xb}, denote by
\be
j(t) := \dim J_{0,t}
\ee
the dimension of the space of wJf of weight 0 and index $t$, and by 
\be
P(t) := \sum_{k=1}^t \left\lceil \frac{k^2}{4t}\right\rceil
\ee
the number of polar terms in the standard region $0\leq k \leq t$. The crucial point is then that for $t>4$ there are more polar terms than wJf,
\be
P(t) > j(t) \qquad t>4\ .
\ee
This means that for a choice of polar coefficients, generically there will not be a corresponding wJf. 
We perform a more systematic analysis of slow growth forms  up to index $t=18$ in the companion paper \cite{Belin:2019rba}.
We show that slow growing SMF are relatively rare, but we seem to always be able to find a slow growing SMF for every wJf of index $t$, with $b = \lfloor \sqrt{t} \rfloor$.

\section*{Acknowledgements}
We are happy to thank Robert Maier for helpful discussions. AB is supported by the NWO VENI grant 680-47-464 / 4114. AC is supported by Nederlandse Organisatie voor Wetenschappelijk Onderzoek (NWO) via a Vidi grant.  This work is supported by the Delta ITP consortium, a program of the Netherlands Organisation for Scientific Research (NWO) that is funded by the Dutch Ministry of Education, Culture and Science (OCW). The work of BM is part of the research programme of the Foundation for Fundamental Research on Matter (FOM), which is financially supported by the Netherlands Organisation for Science Research (NWO).

\appendix

\section{Jacobi forms}\label{app:wjf}

In this appendix we summarise some elementary definitions of Jacobi forms forms used in text, which is based on \cite{MR781735}. 
A Jacobi form, $\varphi_{k,m}(\tau,z)$ of weight $k$ and index $m$ is a holomorphic function on $\H\times\CC\rightarrow\CC$ that has a has Fourier expansion
\be\label{eq:jf3}
\varphi_{k,m}(\tau,z)= \sum_{\substack{n\geq0,l\\ 4mn\geq l^2}} c(n,l) q^n y^l~,\qquad q = e^{2\pi i\tau}\ , \qquad y = e^{2\pi i z}~,
\ee
and its defining transformation properties are 
\be\label{eq:jf1}
\varphi_{k,m}\le({a\tau+b\over c\tau +d},{z\over c\tau +d}\ri)= (c\tau +d)^k\exp\le({2\pi i m c z^2\over c\tau +d}\ri)\varphi_{k,m}(\tau,z)~,\quad  \forall \twobytwo{a}{b}{c}{d} \in SL(2,\ZZ)~,
\ee
and 
\be\label{eq:jf2}
\varphi_{k,m}\le(\tau,{z+ \lambda \tau +\mu}\ri)= \exp\le(-{2\pi i m (\lambda^2\tau+2\lambda z +\mu)}\ri)\varphi_{k,m}(\tau,z)~, \quad \lambda,\mu \in \ZZ~.
\ee
Here $k$ is the {weight} and $m$ is the {index}.  We define the \emph{discriminant} $\Delta:= 4nm-l^2$.
The coefficients $c(n,l)$ then only depend on $\Delta$ and $l$ (mod $2m$),
and in fact only on $\Delta$ if $m$ is prime.

There are several special cases and generalizations 
of Jacobi forms which have to do with the summation range
in (\ref{eq:jf3}). For example, \emph{Jacobi cusp forms} are Jacobi forms
for which $c(0,l)=0$. In particular they vanish at the cusp
$\tau = i\infty$. \emph{Weak Jacobi forms} are holomorphic functions that
satisfy (\ref{eq:jf1}) and (\ref{eq:jf2}), 
but for which we don't impose the condition that $c(n,l)=0$ if 
$\Delta<0$. One can however show that we have $c(n,l)=0$ if
$\Delta < - m^2$, leading to a Fourier expansion
\be
\varphi_{k,m}(\tau,z)= \sum_{\substack{n\geq0,l\\ 4mn- l^2\geq -m^2}} c(n,l) q^n y^l\ .
\ee

A wJf that we regularly used in the paper is
\bea\label{eq:jfex}
\phi_{0,1}&=& 4\le({\theta_2(\tau,z)^2\over \theta_2(\tau)^2}+{\theta_3(\tau,z)^2\over \theta_3(\tau)^2}+{\theta_4(\tau,z)^2\over \theta_4(\tau)^2}\ri)~,
\eea
where $\theta_i(\tau,z)$ are the usual theta functions, and $\theta_i(\tau)\equiv\theta_i(\tau,0)$. And the Jacobi-theta series is
\be
\theta(\tau,z)= -q^{1/8} y^{-1/2} \prod_{n\geq1}(1-q^{n-1}y)(1-q^{n}y)(1-q^{n})~.
\ee

\section{Higher order poles}\label{app:high}

\subsection{General behavior at large $m$ \label{higherordergen}}
In this Appendix we illustrate how to deal with higher order poles. For $\Phi_k$ the exponential lift of a wJf with a pole of order $m_{D,b}>1$ at $p^t=q^{-a}y^{b}$ we obtain from the Residue theorem
\begin{align}\nonumber
&\textrm{Res}\left(\frac{p^{-m}q^{-n}y^{-l}}{\Phi_{k}(p,q,y)}, \bf{p^t=q^{-\textit{a}}y^{b}}\right)= \frac{1}{(m_{D,b}-1)!}\lim_{p\rightarrow q^{-\textit{a}/t}y^{b/t}}\frac{\partial^{m_{D,b}-1}}{\partial p^{m_{D,b}-1}}\left(\left(1-\frac{p^tq^a}{y^b}\right)^{m_{D,b}}\frac{p^{-m}q^{-n}y^{-l}}{\Phi_{k}(p,q,y)}\right)~\\ \label{eq:generalResidue}
&=  \frac{q^{-n}y^{-l}}{(m_{D,b}-1)!}\lim_{p\rightarrow q^{-\textit{a}/t}y^{b/t}}\sum_{s=0}^{m_{D,b}-1}\binom{m_{D,b}-1}{s}\left(p^{-m}\right)^{(s)}\left(\hat{\Phi}_k^{-1}(p,q,y)\right)^{(m_{D,b}-1-s)}~.
\end{align}
In going to the second line we used the Leibniz rule and by $\hat{\Phi}_k$ we denote the part of $\Phi_{k}$ where we stripped off the factor of $(1-p^tq^ay^{-b})^{m_{D,b}}$. 
For large $m$ the term with $s=m_{D,b}-1$ is dominating the residue (\ref{eq:generalResidue}), since for $s<m_{D,b}-1$ at least one derivative acts on $\hat{\Phi}_k$ contributing at most of order  $\mathcal{O}(m^{m_{D,b}-2})$ to the residue. Additionally none of the sums arising in (\ref{eq:generalResidue}) when the derivative acts on $\hat{\Phi}_k$ depend explicitly on $m$, and so they cannot compete with the derivatives acting on $p^{-m}$.
This is illustrated in the example of (\ref{hopexample}).
Consequently we obtain for the leading contribution to (\ref{eq:generalResidue}) 
\begin{align}
&\textrm{Res}\left(\frac{p^{-m}q^{-n}y^{-l}}{\Phi_{k}(p,q,y)}, \bf{p^t=q^{-\textit{a}}y^{b}}\right)~\sim ~(-1)^{m_{D,b}-1}\frac{P_{m_{D,b}-1}(m)}{(m_{D,b}-1)!}\times\frac{q^{-n}y^{-l-m-m_{D,b}+1}}{\hat{\Phi}_k(q^{-\textit{a}/t}y^{b/t},q,y)}~,
\end{align}
where the tilde indicates the approximation in the large $m$ limit and $P_{m_{D,b}-1}(m)$ is a polynomial in $m$ of order $m_{D,b}-1$.

\subsection{An explicit example}\label{hopexample}
We now discuss (\ref{eq:generalResidue}) for a specific example: the reciprocal of the exponential lift of  $4\phi_{0,1}$, which leads to a paramodular form $\Phi_{20}$. More explicitly, we will study 
\begin{align} \label{eq:lift4phi01}
{1\over \Phi_{20}}&=\textrm{Exp-Lift}(-4\phi_{0,1})\cr
&=\frac{1}{q^{2}p^{2}y^{2}}\prod_{\substack{m,n=0\\ l<0}}\frac{1}{(1-y^l)^{c(0,l)}}\prod_{\substack{n>0 \\ l\in \mathbb{Z}}}\frac{1}{(1- q^ny^l)^{c(0,l)}}\prod_{\substack{n, l\in \mathbb{Z} \\ m>0}}\frac{1}{(1- q^ny^lp^{m})^{c(nm,l)}} ~.
\end{align}
This form  has poles characterized by the Humbert surface $H_1(1)$ with $m_{1,1}=4$. The residue at $p=y$ is
\begin{align}
&\textrm{Res}_{\rho=z}\left(\frac{p^{-m}q^{-n}y^{-l}}{\Phi_{20}},\bf{p=y}\right)=\frac{1}{6}\lim_{p\rightarrow y} \frac{\partial^3}{\partial p^3}\left(\left(1-\frac{p}{y}\right)^4\frac{p^{-m}q^{-n}y^{-l}}{\Phi_{20}}\right)~.
\end{align}
In the large $m$ limit the leading contribution for the above residue is obtained when all $p$ derivatives act on the numerator 
\begin{align}
\textrm{Res}_{\rho=z}\left(\frac{p^{-m}q^{-n}y^{-l}}{\Phi_{20}}, \bf{p=y}\right)\sim-\frac{1}{6}m(m+1)(m+2)y^{-m-l-3}q^{-n}\eta(z)^{-48}\eta(\tau-z)^{-48}~.
\end{align}
For this example it is also straight forward to examine the sub leading contributions. 
Since the Humbert surface is $H_1(1)$ , we can map the pole $p=y$ to $y=1$, i.e. we can expand around $z=0$ and examine the residue there. Close to $z=0$, we can directly expand the first part of (\ref{eq:lift4phi01}) and use the fact that $\sum\limits_l c(n,l)=0$ for $n>0$; this gives
\begin{align}\label{eq:m4phi01}
&{\rm Exp-Lift}(-4\phi_{0,1})\approx (2\pi)^{-4}z^{-4} \eta(\rho)^{-48}\eta(\tau)^{-48} \cr
&\times\prod_{\substack{n'>0 \\ l'\in \mathbb{Z}}}\left(1- \frac{q^{n'}}{(1-q^{n'})}\left(e^{2\pi i l' z}-1\right)\right)^{-c(0,l')}\prod_{\substack{m'>0 \\ l',n'\in \mathbb{Z}}}\left(1- \frac{p^{m'}q^{n'}}{(1-p^{m'}q^{n'})}\left(e^{2\pi i l' z}-1\right)\right)^{-c(m'n',l')}
\end{align}
where $\eta(\tau)$ is the Dedekind-eta function. We will now provide some details explaining the sub-leading character of the products in the second line of (\ref{eq:m4phi01}).
Since $c(0,\pm 1)$=4, the expansion of the first product in (\ref{eq:m4phi01}) around  $z=0$  leads to 
\begin{align}\label{eq:mequal0}
\prod_{m'>0}\left(1+4\pi^2 z^2\frac{q^{m'}}{(1-q^{m'})^2}\right)^{-4}=1- 16\pi^2 z^2\sum_{m'=1}^{\infty}\frac{q^{m'}}{(1-q^{m'})^2}~,
\end{align}
while close to $z=0$ we obtain for the second factor 
\begin{align}\label{eq:secondprod}
=&\prod_{l',m'>0}\left(1 + 4\pi^2 l'^2 z^2\frac{p^{m'} q^{n'}}{(1-p^{m'} q^{n'})^2}\right)^{-c(m'n',l')}~.
\end{align}
Now we notice that for a given $m'n'$ we have $\sigma_0(m'n')$ different possiblities to choose $m'$ and $n'$. Additionally observing
\begin{align}\label{eq:relationsigma1}
-\frac{1}{24}\sum_{m'+n'=k}\sum_{l'\geq 0}l'^2 c(k,l')=\sigma_1(k)~.
\end{align}
(\ref{eq:secondprod}) is then equal to 
\begin{align}\label{eq:generalmn}
& 1 + 96\pi^2 z^2\sum_{m',n' >0}\sigma_1(m'n')\sum_{\substack{0<s\leq \sigma_0(m'n')\\ s|(m'n')}}\frac{p^sq^{m'n'/s}}{(1-p^s q^{m'n'/s})^2}~.
\end{align}
Finally combining (\ref{eq:mequal0}) and (\ref{eq:generalmn}), we find the following small $z$ behaviour for  $1/\Phi_{20}$
\begin{align}\nonumber
&\textrm{Exp-Lift}(-4\phi_{0,1})\approx z^{-4}(2\pi i)^{-4}\eta(\rho)^{-48}\eta(\tau)^{-48}\times \\ \label{eq:final4phi01}
&\left(1 + 16\pi^2 z^2\left(6 \sum_{m',n' >0}\sigma_1(m'n')\sum_{\substack{s=0\\ s|(m'n')}}^{\sigma_0(m'n')}\frac{p^sq^{m'n'/s}}{(1-p^s q^{m'n'/s})^2}-\sum_{n'=1}^{\infty}\frac{q^{n'}}{(1-q^{n'})^2}- \sum_{m'=1}^{\infty}\frac{p^{m'}}{(1-p^{m'})^2}\right)\right)~.
\end{align}
The goal now is to extract the Fourier coefficient $d(m,n,l)$ from (\ref{eq:final4phi01}). By mapping back the pole $z=0$ to the pole $z-\rho=0$, we find that
\be \label{dphi20}
d(m,n,l)= -\frac{l^3}{6} \int {d\tau d\rho}\,\frac{q^{-n}p^{-(m+l)}}{\eta(\tau)^{48}\eta(\rho)^{48}} - l  \int d\tau d\rho\,  q^{-n}p^{-(m+l)}F(\tau,\rho) \,,
\ee
where
\begin{align}
F(\tau,\rho)= \frac{4}{\eta(\tau)^{48}\eta(\rho)^{48}}\Big(&6 \sum_{m',n' >0}\sigma_1(m'n')\sum_{\substack{s=0\\ s|(m'n')}}^{\sigma_0(m'n')}\frac{p^sq^{m'n'/s}}{(1-p^s q^{m'n'/s})^2}\cr
&-\sum_{n'=1}^{\infty}\frac{q^{n'}}{(1-q^{n'})^2}- \sum_{m'=1}^{\infty}\frac{p^{m'}}{(1-p^{m'})^2}\Big)~.
\end{align}

From this, we can conclude the following: In the single pole regime  $l\sim -m+\mathcal{O}(1)$, 
and therefore $d(m,n,l)$ scales as $m^3$. The second term in \rref{dphi20} is sub-leading and scales like $m$. As explained more generally at the beginning of Sec.\,\ref{higherordergen}, the growth of the coefficients in $F(\tau,\rho)$ cannot overcome this suppression since we are extracting the $n$ and $m+l$-th powers of $q$ and $p$ respectively, and
\be
n, l+m \sim \mathcal{O}(1) \,.
\ee

For this example, we can however even understand the subleading growth coming from $F(\tau,\rho)$. The divisor function $\sigma_1(s)$ is known to have asymptotic growth
\be
s\log\log s \,,
\ee 
so the growth of the subleading term is still largely dominated by the $\eta$ functions.

\bibliographystyle{JHEP-2}
\bibliography{ref}

\end{document}